\def\rn{}
\def\nn#1 #2{#2. #1}                % Name with 1 initial
\def\nnn#1 #2 #3{#2. #3. #1}            % Name with 2 initials
\def\nnnn#1 #2 #3 #4{#2. #3. #4 #1}     % Name with 3 initials
\def\nnnnn#1 #2 #3 #4 #5{#2. #3. #4 #5. #1} % Name with 4 initials
\def\rf#1;#2;#3;#4;#5 {{\frenchspacing\par\rn#1, #3 {\bf #4}, #5 (#2). \par}}
\def\rrf#1;#2;#3;#4;#5 {{\frenchspacing\rn#1, #3 {\bf #4}, #5 (#2);~}}
\def\rrrf#1;#2;#3;#4;#5 {{\frenchspacing\rn#1, #3 {\bf #4}, #5 (#2).}}
\def\rg#1;#2;#3;#4;#5;#6 {{\frenchspacing\par\rn#1, #3 {\bf #4}, #5 (#2). \par}}
\def\rfbook#1;#2;#3;#4;#5 {{\frenchspacing\par\rn#1, {\it #3} (#5, #4, #2).\par}}
\def\rfprep#1;#2;#3 {{\par\frenchspacing\rn#1, #3 (#2).\par}}
\def\rrfprep#1;#2;#3 {{\frenchspacing\rn#1, #3 (#2);~}}
\def\rrrfprep#1;#2;#3 {{\frenchspacing\rn#1, #3 (#2).}}
\def\rfproc#1;#2;#3;#4;#5;#6 {{\frenchspacing\par\rn#1 #2, in {\it #3}, ed. #4 (#5: #6)\par}}
\def\rfprocp#1;#2;#3;#4;#5;#6;#7 {{\frenchspacing\par\rn#1 #2, in {\it #3}, ed. #4 (#5: #6), p#7\par}}
\def\rg#1;#2;#3;#4;#5;#6 {\par\rn#1 #2, {\it #3}, {\bf #4}, #5 (``#6'') \par}
\def\rf#1;#2;#3;#4;#5 {\par\rn#1, {\it #3}, {\bf #4}, #5 (#2)\par}
\def\rfbook#1;#2;#3;#4;#5 {{\frenchspacing\par\rn#1, {\it #3} (#4: #5, #2)\par}}
\def\rfproc#1;#2;#3;#4;#5;#6 {{\frenchspacing\par\rn#1 #2, in {\it #3}, ed. #4 (#5: #6)\par}}
\def\rfprocp#1;#2;#3;#4;#5;#6;#7 {{\frenchspacing\par\rn#1 #2, in {\it #3}, ed. #4 (#5: #6), p#7\par}}
\def\rfprep#1;#2;#3  {{\par\rn#1, #3, #2\par}}
\def\rfprepp#1;#2;#3 {{\par\rn#1 #2, #3\par}}

\def\etal{{\frenchspacing\it et al.}}

\def\beq#1{\begin{equation}\label{#1}}
\def\eeq{\end{equation}}
\def\beqa#1{\begin{eqnarray}\label{#1}}
\def\eeqa{\end{eqnarray}}

\def\Fig#1{Figure~\ref{#1}}

\def\spose#1{\hbox to 0pt{#1\hss}}
\def\simlt{\mathrel{\spose{\lower 3pt\hbox{$\mathchar"218$}}
     \raise 2.0pt\hbox{$\mathchar"13C$}}}
\def\simgt{\mathrel{\spose{\lower 3pt\hbox{$\mathchar"218$}}
     \raise 2.0pt\hbox{$\mathchar"13E$}}}
\def\simpropto{\mathrel{\spose{\lower 3pt\hbox{$\mathchar"218$}}
     \raise 2.0pt\hbox{$\propto$}}}

\def\ed{\end{document}}

\def\f{X}

\def\Om{\Omega_m}

\def\zcmb{z_{\rm CMB}}
\def\ztwodf{z_{\rm 2df}}

\def\beq#1{\begin{equation}\label{#1}}
\def\eeq{\end{equation}}
\def\beqa#1{\begin{eqnarray}\label{#1}}
\def\eeqa{\end{eqnarray}}

\documentclass[twocolumn,amsmath,nofootinbib]{revtex4} % For astro-ph
\usepackage{graphicx}
\usepackage{url}
\begin{document}
% Include Rokicki's epsf.sty file for Encapsulated PostScript graphics
\input{epsf.sty}

\def\affilmrk#1{$^{#1}$}
\def\affilmk#1#2{$^{#1}$#2;}

\title{Constraints on Oscillating Quintom from Supernova, Microwave Background and Galaxy Clustering}

\author{Jun-Qing Xia, Bo Feng, and Xinmin Zhang }
\affiliation{ Institute of High Energy Physics, Chinese
Academy of Science, P.O. Box 918-4, Beijing 100049, P. R. China}
%\date{\today.}

\begin{abstract}
We consider in this paper a simple oscillating Quintom model of
dark energy which has two free parameters and an equation of state
oscillating and crossing -1. For low redshifts the equation of
state of this model resembles itself similar to the linearly
parameterized dark energy, however differ substantially at large
redshifts. We fit our model to the observational data separately
from the new high redshift supernova observations from the
HST/GOODS program and previous supernova, CMB and galaxy
clustering.
Our results show that because of
the oscillating feature of our model  the constraints from
observations at large redshifts such as CMB become less stringent.

\end{abstract}

\pacs{98.80.Es}

\maketitle

%%%%%%%%%%%%%%%%%%%%%%%%%%%%%%%%%%%%%%%%%%%%%%
%%%%%%%%%%%%%%%%%%%%%%%%%%%%%%%%%%%%%%%%%%%%%%

\setcounter{footnote}{0}

%\sectio{Introduction}

There are strong evidences that the Universe is spatially flat and
accelerating at the present time \cite{Riess04,0407372,Spergel03}.
In the context of Friedmann-Robertson-Walker cosmology, this
acceleration is attributed to the domination of a component,
dubbed dark energy. The simplest candidate for the dark energy
 seems to be a remnant small
   cosmological constant or (true or false) vacuum energy, but it suffers
from the difficulties
associated with the
 fine tuning and coincidence problem, so many
   physicists are attracted with the idea that
the acceleration is driven by dynamical scalar fields, such as
Quintessence\cite{pquint,quint}.

Despite the current theoretical ambiguity for the nature of dark
energy, the prosperous observational data ({\it e.g.} supernova,
CMB and large scale structure data and so on ) have opened a
robust window for testing the recent and even early behavior of
dark energy using some parameterizations for its equation of
state. The recent fits to current supernova(SN) Ia data in the
literature find that even though the behavior of dark energy is to
great extent in consistency with a cosmological constant, an
evolving dark energy with the equation of state $w$ larger than -1
in the recent past but less than -1 today is mildly
favored\cite{running,FWZ}. If such a result holds on with the
accumulation of observational data, this would be a great
challenge to current cosmology.
 Firstly, the cosmological constant as a
candidate for dark energy will be excluded and dark energy must be
dynamical. Secondly, the simple dynamical dark energy models
considered vastly in the literature like the
quintessence\cite{pquint,quint} or the phantom\cite{phantom} can
not be satisfied either.

In the quintessence model, the energy density and the pressure for
the quintessence field are
\begin{equation}
 \rho=\frac{1}{2}\dot Q^2+V(Q)~,~~p=\frac{1}{2}\dot Q^2-V(Q)~.
\end{equation}
 So, its equation of state $w=p/\rho$ is in the range
$-1\leq w\leq 1$ for $V(Q)>0$. However, for the phantom which has
the opposite sign of the kinetic term compared with the
quintessence in the Lagrangian,
\begin{equation}
 \mathcal{L}=-\frac{1}{2}\partial_{\mu}Q\partial^{\mu}Q-V(Q)~,
 \end{equation}
the equation of state $w=(-\frac{1}{2}\dot
Q^2-V)/(-\frac{1}{2}\dot Q^2+V)$ is located in the range of $w\leq
-1$. Neither the quintessence nor the phantom alone can fulfill
the transition from $w>-1$ to $w<-1$ and vice versa.

In Ref.\cite{FWZ} we have proposed a new scenario of the dark
energy, dubbed {\it Quintom}. A simple Quintom model consists of
two scalar fields, one being the quintessence with the other being
the phantom field. This type of Quintom model will provide a
scenario where at early time the quintessence dominates with
$w>-1$ and lately the phantom dominates with $w$ less than $-1$,
satisfying current observations\cite{relev}. A detailed study on
the cosmological evolution of this class of Quintom model is
performed in Ref.\cite{GPZZ}. The Quintom models are different
from the quintessence or phantom in the determination of the
evolution and fate of the universe. Generically speaking, the
phantom model has to be more fine tuned in the early epochs to
serve as dark energy today, since its energy density increases
with expansion of the universe. Meanwhile the Quintom model can
also preserve the tracking behavior of quintessence, where less
fine tuning is needed. In Ref.\cite{FLPZ} we have studied a class
of Quintom models with an oscillating equation of state and found
that oscillating Quintom can unify the early inflation and current
acceleration of the universe, leading to oscillations of the
Hubble constant and a recurring universe. Our oscillating Quintom
would not lead to a big crunch nor big rip. The scale factor keeps
increasing from one period to another and leads naturally to a
highly flat universe. The universe in this model recurs itself and
we are only staying among one of the epochs, in which sense the
coincidence problem is reconciled. In this paper we study the
current observational constraints on the model of oscillating
Quintom. Specifically we consider a variation of the model in
Ref.\cite{FLPZ} with an equation of state given by
\begin{equation}\label{osceq2}
w(z)=w_0+w_1 \sin z ~.
\end{equation}
The model above has two parameters $w_0$ and $w_1$ and for low redshifts
it reduces to
\begin{equation}
 w(z)=w_0+w_1 z ~,
\end{equation}
which is commonly used in the literature for the linear
parameterization of the dark energy. At large redshifts the $w$ of
our model oscillates and will differ substantially from the model
with linear parameterization, which is the key as we will show
below to evade the constraints from the high redshifts
observational data.

  We use the "gold" set of 157 SN Ia  data published by Riess
{\etal} in \cite{Riess04} in fitting to supernova. We assume a
flat space and use a uniform prior on the matter density fraction
of the universe: $0.2<\Omega_m< 0.4$. We constrain the Hubble
parameter to be uniformly in $3\sigma$ HST region: $0.51<h<0.93$.

 In a flat Universe, the luminosity distance can be
expressed as $d_L(z)=(1+z)\Gamma(z)c/H_0$, where
$\Gamma(z)=\int_0^z dz'/E(z')$
 and
\begin{equation}
E(z) \equiv \left[\Omega_m (1+z)^3 +(1-\Omega_m)
\f(z)\right]^{1/2} = H(z)/H_0~,
\end{equation}
 where $X(z)\equiv\rho_X(z)/\rho_X(0)$ and the label of "X" stands for
dark energy. As an illustrative effect we use similar fitting
methods by Refs.\cite{Knop03,wangy04}. For the Cosmic Microwave
Background(CMB) fitting we use here only the CMB shift parameter
\cite{Bond97}, $R\equiv \Om^{1/2}\Gamma(\zcmb)=1.716\pm 0.062$ as
in Ref.\cite{wangy04} where $\zcmb=1089$. The large scale
structure(LSS) information we use is the linear growth rate
$f(\ztwodf)=0.51\pm 0.11$ measured by the 2dF  survey (2dFGRS)
\cite{2dF,Knop03}, where $\ztwodf=0.15$ and $f\equiv (d\ln D/d\ln
a)$, $D\equiv \delta_k(z)/\delta_k(0)$ is  the linear growth
factor and the density perturbation equation of $\delta_k$
evolutes as
 $\ddot{\delta}_k + 2 (\dot{a}/a)\dot{\delta}_k - 4 \pi G \rho_{\rm
M} \delta_k=0$.

\begin{figure}[htbp]
\begin{center}
\includegraphics[scale=0.6]{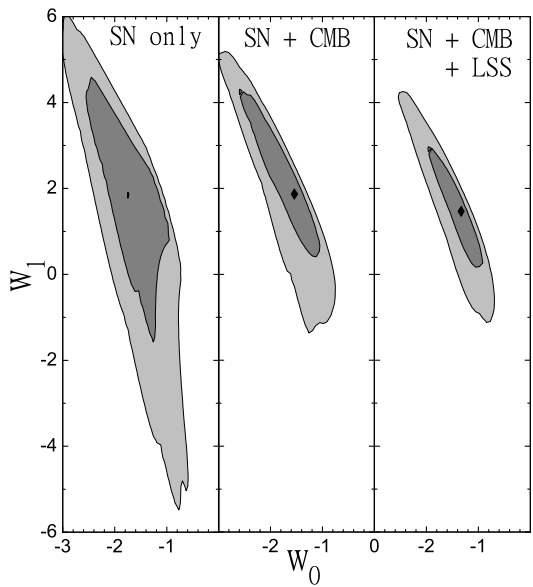}
\caption{SN Ia, CMB and LSS constraints on the oscillating dark
energy model.  The best fit values are shown in the centers of
each panel. The grey and light grey areas show the $1\sigma$ and
$2\sigma$ confidence regions respectively. \label{fig:osc}}
\end{center}
\end{figure}

\begin{figure}[htbp]
\begin{center}
\includegraphics[scale=0.5]{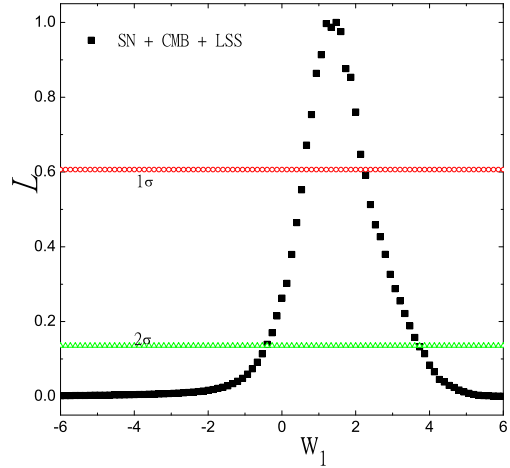}
\caption{1-dimensional likelihood of the variation amplitude of
$w_1$ using SN Ia, CMB and LSS, shown with the $1\sigma$ and
$2\sigma$ limits. \label{fig:1d}}
\end{center}
\end{figure}

\begin{figure}[htbp]
\begin{center}
\includegraphics[scale=0.5]{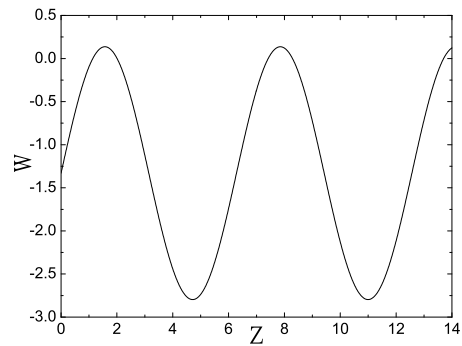}
\caption{An illustration of the equation of state with redshift
for best fit value of \Fig{fig:osc} using SN Ia $+$ CMB $+$ LSS,
where $(w_0,w_1)=(-1.33,1.47)$. \label{fig:will}}
\end{center}
\end{figure}

\begin{figure}[htbp]
\begin{center}
\includegraphics[scale=0.5]{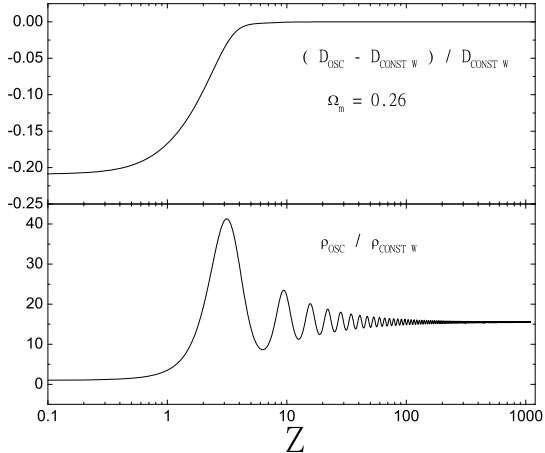}
\caption{A comparison of linear growth factor and density between
oscillating dark energy $w=-1.33 + 1.47 \sin z$ and that with a
constant equation of state $w=-1.33 $. \label{fig:ill}}
\end{center}
\end{figure}

\begin{figure}[htbp]
\begin{center}
\includegraphics[scale=0.5]{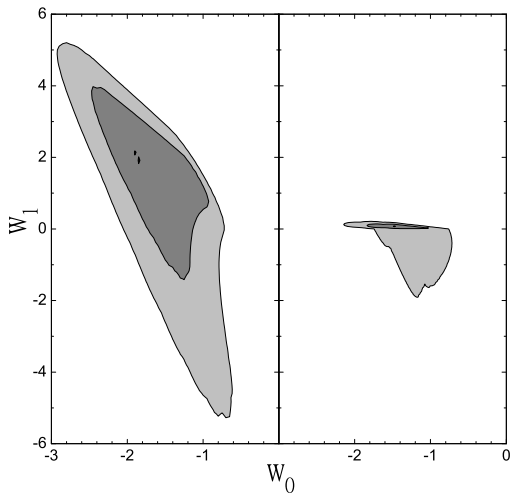}
\caption{SN Ia and CMB  constraints on the linear parameterized
dark energy model. Left : SN Ia only; Right: SN Ia $+$ CMB
constraints. \label{fig:lin}}
\end{center}
\end{figure}

\begin{figure}[htbp]
\begin{center}
\includegraphics[scale=0.5]{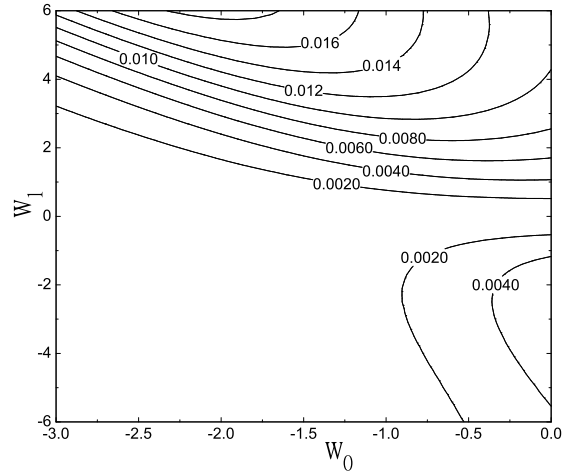}
\caption{A comparison of the maximum luminosity distance
difference up to $z=1.$ for oscillating dark energy and the linear
parameterized dark energy model. The contour stands for $|\Delta
d_L/d_L|$. \label{fig:ill0}}
\end{center}
\end{figure}

In \Fig{fig:osc} we delineate SN Ia, CMB and LSS constraints on
the oscillating dark energy model. We can find a large parameter
space is allowed by current supernova data while additional
constraints from CMB $R$ parameter and linear growth rate of 2df
have restricted considerably the parameters within a lower region.
 We plot the  corresponding  one dimensional
likelihood of $w_1$ in \Fig{fig:1d}.
 An oscillating Quintom model seems to be
mildly favored at 1$\sigma$. The best fit value for the SN Ia $+$
CMB $+$ LSS of our oscillating model is $(w_0,w_1)=(-1.33,1.47)$.
In \Fig{fig:will} we show the corresponding $w(z)$.  As $R$
contains the contribution of $\rho_X$ on all relevant scales as an
integration, $w_1$ is somewhat constrained, this has also
reflected the weak correlation of our parameterization. When LSS
constraint is also taken into account we find although the shape
of the parameter space is not changed, the best fit parameter has
a lower $w_1$ compared with SN Ia and CMB constraints. In the
upper panel of \Fig{fig:ill} we find that the linear growth factor
deviates reasonably from a constant $w_0$ case in low redshift. We
should point out in our simple two parameter expansion the period
of oscillation is fixed, which is due to the simplification of our
parameterization.

Before concluding we compare our model in (3) with the linearly
parameterized dark energy model in (4). In \Fig{fig:lin} we show
the SN Ia and CMB constraints on linear parameterized dark energy
model in (4). Comparing \Fig{fig:lin} and \Fig{fig:osc} one can
see the similarity for the linear and oscillating models when
considering only the supernova data. This is understandable since
our model in (3) for low redshifts reduces to the one in (4),
however when including the high redshift observational data such
as CMB the difference is significant. To understand these firstly
we show in \Fig{fig:ill0} the maximum difference of $\Delta
d_L/d_L$ of the two dark energy models up to $z<1.$, where only 9
SN Ia of the gold sample lie outside the range, the maximum
difference is no more than $2\%$. In the calculations we have set
$\Omega_m=0.26$ and $h=0.735$ as an example. The maximum
difference is considerably small and the parameter spaces are
expected to be similar for the two models under supernova
constraints.
 Meanwhile as from \Fig{fig:lin} we
also expect $w_0\lesssim -0.5$ for the oscillating case, in the
early epoch as
\begin{equation}
\f=a^{-3(1+w_0)} e^{-3 w_1 \int^a_1 \frac{\sin(a'^{-1}-1)}{a'}
da'} ~,
\end{equation}
the contribution of $w_1$ term to $\rho_X$ would be cancelled due
to averaging over the oscillating periods, as can be seen from
\Fig{fig:ill}. In \Fig{fig:ill} we have shown the ratio of density
with $w(z)=w_0+w_1 \sin z$ and  $w(z)=w_0$, assuming $w_0=-1.33$
and $w_1=1.47$ and the same background parameters as
\Fig{fig:ill0}. We can see the ratio is a constant in the early
epochs. Dark energy would hence be negligible on large redshifts
compared with matter energy density and
 the $R$ parameter of CMB would not give very strong constraints on
the oscillating dark energy. When CMB is considered in the linear
parameterization of the equation of state the upper limit shrinks
promptly to near zero. This is because that $\rho_X(z)/\rho_X(0)$
would blow up for a sizable positive $w_1$ on large redshifts and
hence CMB has restricted the model parameter $w_1$  much stronger
than supernova.

In summary we in this paper have considered a simple two-parameter
model of oscillating Quintom. The  oscillating equation of state
resemble itself similar to the linear parameterized dark energy in
low redshift, meanwhile the oscillating feature of the model make
it possible to evade the constraints from CMB observations. Our
results indicate that a dynamical model of dark energy such as the
oscillating Quintom proposed in this paper is mildly favored under
current observations. Our analysis here can be generalized into
different Quintom model with an oscillating equation of state as a
function of scale factor $a$ for example $w(a)=w_0 + w_1
\sin(1-a)$ which becomes the one proposed by
Refs.\cite{DP,linder,0407372} for small $(1-a)$.

{\bf Acknowledgements:} We thank  Alexei Starobinsky, Constantinos
Skordis, Yun Wang, Jun'ichi  Yokoyama  and Zhong-Hong Zhu  for
helpful discussions. We thank   Xiulian Wang, Xiaojun Bi, Peihong
Gu, Hong Li and Yunsong Piao for useful conversations. This work
is supported in part by National Natural Science Foundation of
China under Grant Nos. 90303004 and 19925523 and by Ministry of
Science and Technology of China under Grant No. NKBRSF G19990754.

 %\newpage

\end{document}